\documentclass[12pt]{article}  
\usepackage{amsmath}  
\usepackage{graphicx}
\usepackage[latin1]{inputenc}

\def\Nab{{\mbox{\boldmath $\nabla$}}}
\def\fig{Fig.\;\ref}
\def\graic{\center \includegraphics[height=0.2\textheight, width=1.0\textwidth]}

\begin{document}

\title{A discrete approach to the vacuum Maxwell equations and the fine structure constant}

\author{Wolfgang Orthuber\\
\footnotesize Klinik für Kieferorthopädie der Universität Kiel \\*[-0.2cm]
\footnotesize Arnold-Heller-Str. 16, D-24105 Kiel, Germany \\*[-0.2cm]
\footnotesize further info: http://www.orthuber.com   }

\date{}

\maketitle 

\begin{abstract}
\hspace{-2em}
We recommended consequent discrete combinatorial research in mathematical physics \cite{or2}. Here we show an example how discretization of partial differential equations can be done and that quickly unexpected new findings can result from research in this up to now unexplored area. We transformed the vacuum Maxwell equations into finite-difference equations, provided simple initial conditions and studied the development of the electromagnetic fields using special software \cite{or4} \cite{or}. The development is wave-like as expected. But it is not trivial, the wave maxima have different heights. If all (by definition minimal) finite differences of the location coordinates are multiplied by numbers (coupling factors) whose squares are equal to the fine structure constant, we noticed:\\
1. The first two wave maxima have nearly the same height. Of course this can be also coincidental.\\
2. The following maxima are at first slightly decreasing and then, beginning with the $6$th maximum, exponentially increasing.
\end{abstract}
\vspace*{1cm} 

\newpage

\setcounter{tocdepth}{4}
\tableofcontents
\listoffigures
\newpage

\section{Introduction}
We have shown \cite{or1} that only an a priori finite mathematical model can have an exact equivalent in physical reality. This means that it implies an only finite number of basic arithmetic operations on an a priori finite numerical space\footnote{Nevertheless both can increase without boundary \emph{when} time increases without boundary (infinite potential).} which can be represented without using irrational numbers.\\ \\
Up to now there is not much experience in this area: Important physical equations are defined on continuous (a priori infinite) sets and often written as partial differential equations. If we want to find the natural finite basis of them, first we have to replace differential calculus by finite-difference calculus. This can soon lead to difficult combinatorics, especially in case of interactions across several dimensions. But increasing performance of computers offers new possibilities. The mentioned numerical space can be represented by finite dimensional numerical lattices (finite dimensional point lattices with numbers resp. \emph{quantities} assigned to every point) which can be handled adequately by a computer. So we developed as help software for handling numerical lattices and for studying the results of numerical algorithms on them. All lattice points are addressed by integer coordinates and the quantities assigned to the points can be complex\footnote{Both complex rational and complex floating-point numbers are supported. Subsequently by the term \emph{lattice} always this kind of numerical lattice is meant.}. We used the software to study the vacuum Maxwell equations. \bigskip 

We converted these into finite-difference equations as directly as possible because we wanted to operate on an elementary level (a priori finite). So we had to avoid analytic methods (like usage of infinite power series, trigonometric functions, complex and matrix exponential functions, Fourier transforms) which are often used to \emph{force} an easy to survey, stable development. 

\section{Material and Method}
\subsection{Software}
\label{subsecSoftware}
The in C++ written open-ended software \cite{or4} \cite{or} is designed for editing of numerical lattices and for generation of statistics like sums over the absolute or squared quantities of subspaces, with graphical representation. Using the ASCII file format of saved data as interface or writing own code the user can check the results of own algorithms. Additionally it is possible to define algorithms without writing code. This is done by usage of a table with several entries. Every entry defines a coupling between two lattices in form of an addition from the first (the source lattice) to the second (the destination lattice). The strength of the coupling is determined by the \emph{coupling factor} $p$: During every iteration of the algorithm all non-zero quantities of the source lattice with maximal iteration number\footnote{The results of every iteration are indexed by an increasing iteration number $n$. This can be used to analyze the temporal development.} are added to shifted locations within the destination lattice after multiplication by $p$.\\
Source and destination lattices, the multidimensional relative shift resp. offset are selectable using integer numbers, the coupling factor $p$ can be a complex number, if desired. The user determines how many iterations of all additions (defined by the table) are done.
\smallskip

\scriptsize 
The motivation for this scheme arises from the fact that it is simple and nevertheless very flexible. Among others it can be used to study random walks like those discussed in \cite{or2}. Note that $p$ also can be a complex probability amplitude and is not restricted to the interval $[0,1]$.

\normalsize
If the number of iterations and additions per iteration is great, the lattice(s) may occupy too much memory of the computer. Therefore it is possible to activate automatic deletion of the lattice points with smallest absolute quantities if these would occupy too much memory. Here we used this after we have checked that up to $150$ iterations in our application it has not affected the $12$ most significant digits of the evaluated numerical data.

\subsection{The Maxwell equations}
Using MKS units, the Maxwell equations in linear and isotropic media are
\begin{eqnarray}
\Nab \cdot \mathbf{E}&=&\frac{\rho}{\epsilon}\ ,\label {eqdiv1E}\\
\Nab \cdot \mathbf{B}&=&0\label {eqdiv1B}\ ,\\
\mu \mathbf{J} + \mu\epsilon\frac{\partial \mathbf{E}}{\partial t}&=&\ \ \Nab \times \mathbf{B} \label {eqrot1B}\ ,\\
\frac{\partial \mathbf{B}}{\partial t}&=&-\Nab \times \mathbf{E}\ ,\label {eqrot1E}
\end{eqnarray}
where $\mathbf{E}$ is the electric field, $\mathbf{B}$ is the magnetic field, $\rho$ is the charge density, $\mathbf{J}$ is the vector current density, $\epsilon$ is the permittivity and $\mu$ is the permeability. In the vacuum which is free of current and electric charge ($\mathbf{J}=0$ and $\rho=0$) they reduce to the \emph{vacuum Maxwell equations} in which besides (\ref{eqdiv1B}) also (\ref{eqdiv1E}) vanishes and (\ref{eqrot1B}), (\ref{eqrot1E}) become
\begin{eqnarray}
\frac{\partial \mathbf{E}}{\partial t}&=&\frac{1}{\mu_0\epsilon_0} \ \Nab \times \mathbf{B} = c^2 \ \Nab \times \mathbf{B}\ ,\notag\\
\frac{\partial \mathbf{B}}{\partial t}&=&-\Nab \times \mathbf{E}\ ,\notag
\end{eqnarray}
where $\epsilon_0$ is the permittivity of free space, $\mu_0$ is the permeability of free space and c is the speed of light in the vacuum. Using the definition
$$
\mathbf{\tilde B}:=c\mathbf{B}
$$
we obtain
\begin{eqnarray}
\frac{\partial \mathbf{E}}{\partial t}&=\ \ c& \Nab \times \mathbf{\tilde B}\ ,\label {eqrot3B}\\
\frac{\partial {\mathbf {\tilde B}}}{\partial t}&=\ -c& \Nab \times \mathbf{E}\label {eqrot3E}\ .
\end{eqnarray}
The proportionality factor $c$ still has the dimension $ms^{-1}$. We will see (cf. \ref{chapAlgorithm}) that during discretization automatically the question about the natural dimensionless relationship arises.
\label{chapMaxwDimensionless} 

\subsection{Discretization}
Let $x,y,z$ denote Cartesian coordinates and let $E_x,E_y,E_z$ denote the components of $\mathbf{E}=\mathbf{E}(t,x,y,z)$ and $\tilde B_x,\tilde B_y,\tilde B_z$ those of $\mathbf{\tilde B}=\mathbf{\tilde B}(t,x,y,z)$. With this (\ref{eqrot3B}) and (\ref{eqrot3E}) are equivalent to 
\footnotesize
\begin{equation}
\label{eqrotBcomp}
\frac{\partial E_x}{\partial t}=\ c \left( \frac{\partial \tilde B_z}{\partial y} - \frac{\partial \tilde B_y}{\partial z} \right),\ 
\frac{\partial E_y}{\partial t}=\ c \left( \frac{\partial \tilde B_x}{\partial z} - \frac{\partial \tilde B_z}{\partial x} \right),\ 
\frac{\partial E_z}{\partial t}=\ c \left( \frac{\partial \tilde B_y}{\partial x} - \frac{\partial \tilde B_x}{\partial y} \right),
\end{equation}
\begin{equation}
\label{eqrotEcomp}
\frac{\partial \tilde B_x}{\partial t}=-c \left( \frac{\partial E_z}{\partial y} - \frac{\partial E_y}{\partial z} \right),
\frac{\partial \tilde B_y}{\partial t}=-c \left( \frac{\partial E_x}{\partial z} - \frac{\partial E_z}{\partial x} \right),
\frac{\partial \tilde B_z}{\partial t}=-c \left( \frac{\partial E_y}{\partial x} - \frac{\partial E_x}{\partial y} \right).
\end{equation}
\normalsize
A careful consideration, however, must take into account that the differences in time and location cannot be arbitrarily small.

\subsubsection{(Half) minimal differences as units}
\label{chapmindiffasunits}
We know that there are non-vanishing minimal absolute differences $dt$ in time and $ds$ in location\footnote{They depend on further variables like energy per photon which influence both $dt$ and $ds$ (but not necessarily in all those cases the quotient $dt/ds$).} so that absolute differences of $t$ below $dt$ and of $x,y,z$ below $ds$ are not existing in physical reality. Therefore it is appropriate to base the scaling on $dt$ and $ds$. We will use $dt$ as unit of time and (because we want to form minimal \emph{symmetric} differences in location) $ds/2$ as unit of location.

\subsubsection{Integer representation}
Define
\begin{equation}
\label{eqdefhattxyz}
\hat t:=\frac{t}{dt},\ \hat x:=\frac{2x}{ds},\ \hat y:=\frac{2y}{ds},\ \hat z:=\frac{2z}{ds}\ \ \ \ \text{and}
\end{equation}
\begin{equation}
\label{eqdefhatEB}
\mathbf{\hat E}(\hat t, \hat x, \hat y, \hat z):=\mathbf{E}(t,x,y,z), \ \mathbf{\hat B}(\hat t, \hat x, \hat y, \hat z):=\mathbf{\tilde B}(t,x,y,z)\ .
\end{equation}
The coordinates $\hat t,\hat x,\hat y,\hat z$ are all dimensionless having only integer differences. Now it is possible to form minimal symmetric finite differences using only integer coordinates. For $v \in \{x,y,z\}$ let $\hat E_v$, $\hat B_v$, $E_v$ and $\tilde B_v$ denote the components of $\mathbf{\hat E}$, $\mathbf{\hat B}$, $\mathbf{E}$ and $\mathbf{\tilde B}$ respectively. Define
\footnotesize
\begin{eqnarray*}
\hat E_{vt}&:=&\hat E_v(\hat t+1,\hat x, \hat y, \hat z)-\hat E_v(\hat t,\hat x, \hat y, \hat z)= E_v(t+dt,x, y, z)- E_v(t,x,y,z)\ ,\\
\hat E_{vx}&:=&\hat E_v(\hat t,\hat x+1, \hat y, \hat z)-\hat E_v(\hat t,\hat x-1, \hat y, \hat z)= E_v(t,x+ds/2, y, z)- E_v(t,x-ds/2,y,z)\ ,\\
\hat E_{vy}&:=&\hat E_v(\hat t,\hat x, \hat y+1, \hat z)-\hat E_v(\hat t,\hat x, \hat y-1, \hat z)= E_v(t,x, y+ds/2, z)- E_v(t,x,y-ds/2,z)\ ,\\
\hat E_{vz}&:=&\hat E_v(\hat t,\hat x, \hat y, \hat z+1)-\hat E_v(\hat t,\hat x, \hat y, \hat z-1)= E_v(t,x, y, z+ds/2)- E_v(t,x,y,z-ds/2)\ ,
\end{eqnarray*}
\begin{eqnarray*}
\hat B_{vt}&:=&\hat B_v(\hat t+1,\hat x, \hat y, \hat z)-\hat B_v(\hat t,\hat x, \hat y, \hat z)=\tilde B_v(t+dt,x, y, z)-\tilde B_v(t,x,y,z)\ ,\\
\hat B_{vx}&:=&\hat B_v(\hat t,\hat x+1, \hat y, \hat z)-\hat B_v(\hat t,\hat x-1, \hat y, \hat z)=\tilde B_v(t,x+ds/2, y, z)-\tilde B_v(t,x-ds/2,y,z)\ ,\\
\hat B_{vy}&:=&\hat B_v(\hat t,\hat x, \hat y+1, \hat z)-\hat B_v(\hat t,\hat x, \hat y-1, \hat z)=\tilde B_v(t,x, y+ds/2, z)-\tilde B_v(t,x,y-ds/2,z)\ ,\\
\hat B_{vz}&:=&\hat B_v(\hat t,\hat x, \hat y, \hat z+1)-\hat B_v(\hat t,\hat x, \hat y, \hat z-1)=\tilde B_v(t,x, y, z+ds/2)-\tilde B_v(t,x,y,z-ds/2)\ ,
\end{eqnarray*}
\normalsize
and transformation of (\ref{eqrotBcomp}) and (\ref{eqrotEcomp}) into finite-difference equations yields
\footnotesize
\begin{equation}
\label{eqEfinitediff1}
	\frac{\hat E_{xt}}{dt}=\ c \left(\frac{\hat B_{zy}}{ds} - \frac{\hat B_{yz}}{ds}\right),\ 
	\frac{\hat E_{yt}}{dt}=\ c \left(\frac{\hat B_{xz}}{ds} - \frac{\hat B_{zx}}{ds}\right),\ 
	\frac{\hat E_{zt}}{dt}=\ c \left(\frac{\hat B_{yx}}{ds} - \frac{\hat B_{xy}}{ds}\right), 
\end{equation}
\begin{equation}
\label{eqBfinitediff1}
	\frac{\hat B_{xt}}{dt}=-c \left(\frac{\hat E_{zy}}{ds} - \frac{\hat E_{yz}}{ds}\right),\ 
	\frac{\hat B_{yt}}{dt}=-c \left(\frac{\hat E_{xz}}{ds} - \frac{\hat E_{zx}}{ds}\right),\ 
	\frac{\hat B_{zt}}{dt}=-c \left(\frac{\hat E_{yx}}{ds} - \frac{\hat E_{xy}}{ds}\right)\ . 
\end{equation}
\normalsize
This would be equivalent to (\ref{eqrotBcomp}) and (\ref{eqrotEcomp}) in the borderline case $dt\rightarrow 0$ and $ds\rightarrow 0$. But due to \ref{chapmindiffasunits} the equations (\ref{eqEfinitediff1}) and (\ref{eqBfinitediff1}) are closer to reality.\\
Using
\begin{equation}
\label{eqDefp}
	\hat{p}:=c\  \frac{dt}{ds}
\end{equation}
as absolute coupling factor they become more compact:
\footnotesize
\begin{eqnarray}
\label{eqEdt}
 \hat E_{xt}&=&\ \ \hat{p} \left(\hat B_{zy} - \hat B_{yz}\right),\  
 \hat E_{yt}=\ \ \hat{p} \left(\hat B_{xz} - \hat B_{zx}\right),\  
 \hat E_{zt}=\ \ \hat{p} \left(\hat B_{yx} - \hat B_{xy}\right),\\ 
\label{eqBdt}
 \hat B_{xt}&=& -\hat{p} \left(\hat E_{zy} - \hat E_{yz}\right),\ 
 \hat B_{yt}= -\hat{p} \left(\hat E_{xz} - \hat E_{zx}\right),\ 
 \hat B_{zt}= -\hat{p} \left(\hat E_{yx} - \hat E_{xy}\right)\ .
\end{eqnarray}
\normalsize

\subsection{The algorithm}
\label{chapAlgorithm}
Written out explicitly from (\ref{eqEdt}) and (\ref{eqBdt}) follows
\footnotesize
\begin{eqnarray}
\hat E_x(\hat t+1,\hat x,\hat y,\hat z)=\hat E_x(\hat t,\hat x,\hat y,\hat z)
&+&\hat{p}\left(\hat B_z(\hat t,\hat x, \hat y+1, \hat z)-\hat B_z(\hat t,\hat x, \hat y-1, \hat z)\right)\notag\\
&-&\hat{p}\left(\hat B_y(\hat t,\hat x, \hat y, \hat z+1)-\hat B_y(\hat t,\hat x, \hat y, \hat z-1)\right)\ ,\notag\\
\hat E_y(\hat t+1,\hat x,\hat y,\hat z)=\hat E_y(\hat t,\hat x,\hat y,\hat z)
&+&\hat{p}\left(\hat B_x(\hat t,\hat x, \hat y, \hat z+1)-\hat B_x(\hat t,\hat x, \hat y, \hat z-1)\right)\notag\\
&-&\hat{p}\left(\hat B_z(\hat t,\hat x+1, \hat y, \hat z)-\hat B_z(\hat t,\hat x-1, \hat y, \hat z)\right)\notag\ ,\\
\hat E_z(\hat t+1,\hat x,\hat y,\hat z)=\hat E_z(\hat t,\hat x,\hat y,\hat z)
&+&\hat{p}\left(\hat B_y(\hat t,\hat x+1, \hat y, \hat z)-\hat B_y(\hat t,\hat x-1, \hat y, \hat z)\right)\notag\\
&-&\hat{p}\left(\hat B_x(\hat t,\hat x, \hat y+1, \hat z)-\hat B_x(\hat t,\hat x, \hat y-1, \hat z)\right)\notag\ ,\\ \label{eqAlgExpplicit}\\
\hat B_x(\hat t+1,\hat x,\hat y,\hat z)=\hat B_x(\hat t,\hat x,\hat y,\hat z)
&-&\hat{p}\left(\hat E_z(\hat t,\hat x, \hat y+1, \hat z)-\hat E_z(\hat t,\hat x, \hat y-1, \hat z)\right)\notag\\
&+&\hat{p}\left(\hat E_y(\hat t,\hat x, \hat y, \hat z+1)-\hat E_y(\hat t,\hat x, \hat y, \hat z-1)\right)\ ,\notag\\
\hat B_y(\hat t+1,\hat x,\hat y,\hat z)=\hat B_y(\hat t,\hat x,\hat y,\hat z)
&-&\hat{p}\left(\hat E_x(\hat t,\hat x, \hat y, \hat z+1)-\hat E_x(\hat t,\hat x, \hat y, \hat z-1)\right)\notag\\
&+&\hat{p}\left(\hat E_z(\hat t,\hat x+1, \hat y, \hat z)-\hat E_z(\hat t,\hat x-1, \hat y, \hat z)\right)\notag\ ,\\
\hat B_z(\hat t+1,\hat x,\hat y,\hat z)=\hat B_z(\hat t,\hat x,\hat y,\hat z)
&-&\hat{p}\left(\hat E_y(\hat t,\hat x+1, \hat y, \hat z)-\hat E_y(\hat t,\hat x-1, \hat y, \hat z)\right)\notag\\
&+&\hat{p}\left(\hat E_x(\hat t,\hat x, \hat y+1, \hat z)-\hat E_x(\hat t,\hat x, \hat y-1, \hat z)\right)\notag\ .
\end{eqnarray}
\normalsize
For stepwise calculation of the temporal development we can, as shown in \cite{or4}, directly implement these equations in our software using 6 four-dimensional numerical lattices, 3 representing $\mathbf{\hat E}$, 3 representing $\mathbf{\hat B}$, and $\pm\hat{p}$ as coupling factor $p$ (cf. \ref{subsecSoftware}). At this $\hat{p}$ is a free variable, so we have to search for reasonable values of it (i.e. automatically we have to search for the natural dimensionless relationship, as mentioned in \ref{chapMaxwDimensionless}). The time steps are uniform because
both laws (\ref{eqrot3B}) and (\ref{eqrot3E}) are considered to be valid at every time, i.e. $\mathbf{\hat E}$ and $\mathbf{\hat B}$ are calculated simultaneously in  (\ref{eqAlgExpplicit}), not alternating. Therefore this algorithm differs from leapfrog FDTD schemes \cite{yee}.\\ \\
The above equations (\ref{eqAlgExpplicit}) show that a linear superposition of the initial conditions leads to a linear superposition of the results. Therefore we chose for the purpose of clarity as simple as possible nontrivial (non-zero) initial conditions. As initial time we defined $\hat t=0$ and set all initial quantities (of $\hat E_x,\hat E_y, \hat E_z, \hat B_x, \hat B_y, \hat B_z$) to zero  on all locations except $\hat E_x$ in the origin $(0,0,0,0)$ which we set to one: $\hat E_x(0,0,0,0):=1$. Then we analyzed the temporal development of $\hat E_x$ resulting from iterative calculation of (\ref{eqAlgExpplicit}) and its dependence of $\hat{p}$ which by definition (\ref{eqDefp}) is a positive dimensionless number.

\section{Results}
As expected a wavelike development results. But it is not trivial, the heights of the wave maxima are not constant. In case of $\hat{p}>1/8$ they are increasing from the beginning, which is not realistic. Therefore we focused our interest on smaller values of $\hat{p}$.
\subsection{Preliminary note}
\label{chapCarefulInterpretation}
Careful evaluation is necessary because (\ref{eqAlgExpplicit}) bases on a simplified model, particularly it is derived from the idealized vacuum situation. Even small errors in (\ref{eqAlgExpplicit}) can become relevant, if many iterations are done. Therefore the calculation of the first wave is more reliable than calculations of other waves and a detailed graphical representation of the first wave is reasonable.

\subsection{Short term development: The first wave}
\label{chapShortTermDevelopment}
\begin{figure}[htb]
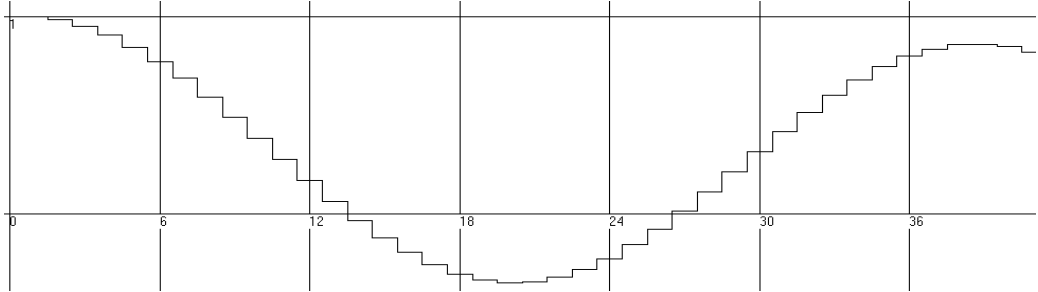
 \graic 
{wqpmf1} \caption[The first wave in case of ${\hat{p}=1/16}$]{The first wave in case of ${\hat{p}=1/16}$. Abscissa${: t}$, \ \ Ordinate${: E_x(t,0,0,0)}$.}
	\label{fig_wqpmf1}
\end{figure}

\begin{figure}[htb]
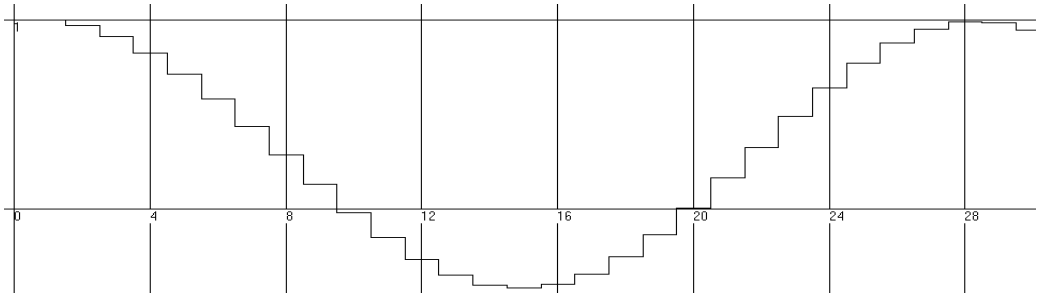
 \graic 
{wqpmf2}	\caption[The first wave in case of ${\hat{p}=\sqrt{\alpha}=0.085424542921}$]{The first wave in case of ${\hat{p}=\sqrt{\alpha}=0.085424542921}$. Abscissa${: t}$, \ \ Ordinate${: E_x(t,0,0,0)}$.}
	\label{fig_wqpmf2}
\end{figure}

\begin{figure}[htb]
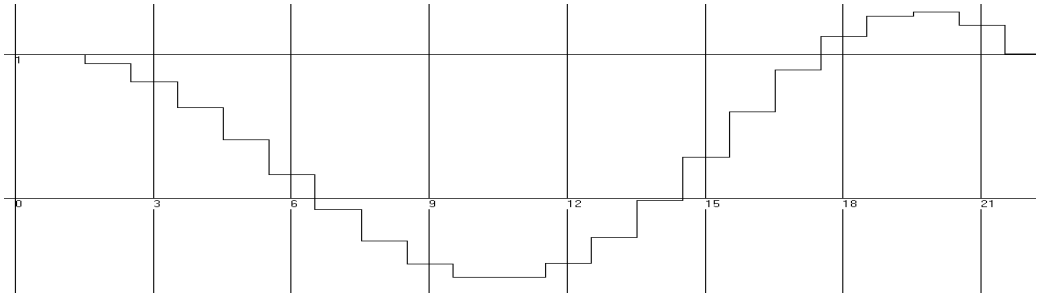
 \graic 
{wqpmf3}	\caption[The first wave in case of ${\hat{p}=1/8}$]{The first wave in case of ${\hat{p}=1/8}$. Abscissa${: t}$, \ \ Ordinate${: E_x(t,0,0,0)}$.}
	\label{fig_wqpmf3}
\end{figure}

We are searching  for realistic values of $\hat{p}$. If these should cause an as stable as possible wave pattern, the wave maxima should have  nearly equal height. Especially the first two maxima (the maxima of the first wave which we can calculate with the best reliability) should have similar heights.
\fig{fig_wqpmf1}, \ref{fig_wqpmf2} and \ref{fig_wqpmf3} show the first wave for three different values of $\hat{p}$. 

\clearpage

We see in \fig{fig_wqpmf1} (resp. \ref{fig_wqpmf3}) that in case of $\hat{p}=1/16$ (resp. $\hat{p}=1/8$) the second maximum is considerably smaller (resp. greater) than the first. More interesting is  \fig{fig_wqpmf2}. There 
$$\hat{p}=\sqrt{\alpha}=0.085424542921=1/11.7062376432\ \ \,$$
in which $\alpha=1/137.03599976$ is the fine structure constant\footnote{It can be derived very precisely from basic experiments, e.g. from measurements of the Quantum Hall Effect \cite{yo}.} as listened in \cite{co}.\\
The calculated quantities near the maximum at $t=28$ are
\begin{eqnarray}
\hat E_x(26,0,0,0)&=&0.877636902081288\ ,\notag\\
\hat E_x(27,0,0,0)&=&0.950716819197347\ ,\notag\\	
\hat E_x(28,0,0,0)&=&0.98752930561647	\ ,\notag\\
\hat E_x(29,0,0,0)&=&0.986271782354442\ ,\notag\\	
\hat E_x(30,0,0,0)&=&0.947405056005354\ .\notag\	
\end{eqnarray}
So in case of $\hat{p}=\sqrt{\alpha}$ the second maximum deviates less than $2\%$ from the first maximum which lies in the start and has the initial quantity $1$.\ \footnote{In case of $\hat{p}=0.08672495$ the maximum also is at $t=28$ and  differs less than $1$ ppm from $1$. But due to \ref{chapCarefulInterpretation} there is no good foundation for such fine interpolation.}

\subsection{Long term development}
\begin{figure}[htb]
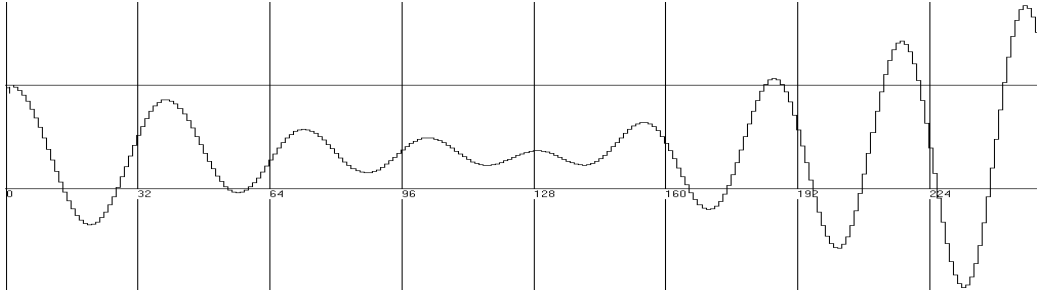
 \graic 
{wqpmf4}	\caption[Long term development in case of ${\hat{p}=1/16}$]{Long term development in case of ${\hat{p}=1/16}$. Abscissa${: t}$, \ \ Ordinate${: E_x(t,0,0,0)}$.}
	\label{fig_wqpmf4}
\end{figure}

\begin{figure}[htb]
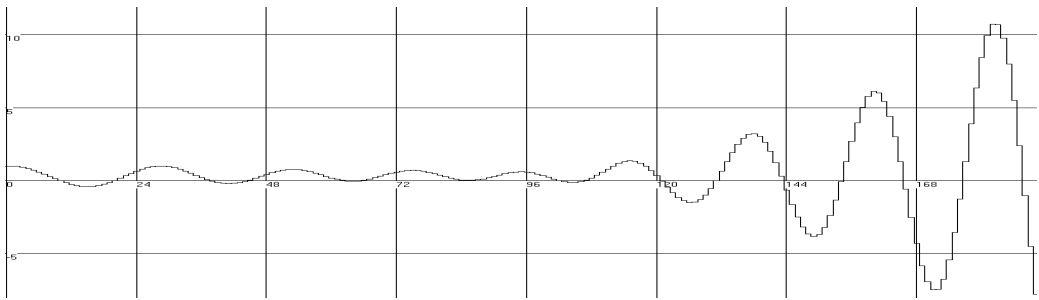
 \graic 
{wqpmf5} \caption[Long term development in case of ${\hat{p}=\sqrt{\alpha}}$]{Long term development in case of ${\hat{p}=\sqrt{\alpha}}$. Abscissa${: t}$, \ \ Ordinate${: E_x(t,0,0,0)}$.}
	\label{fig_wqpmf5}
\end{figure}

\begin{figure}[htb] 
		\center \includegraphics[width=0.7\textwidth]
{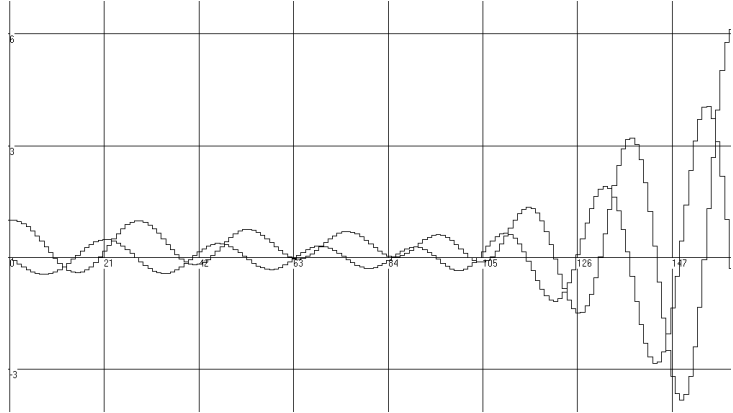} \caption[$E_x$ together with $B_z$]{$E_x$ together with $B_z$ in case of ${\hat{p}=\sqrt{\alpha}}$. Abscissa${: t}$, \ \ Ordinate${: E_x(t,0,0,0)}$ and $B_z(t,0,1,0)$.}
	\label{fig_wqpmf7}
\end{figure}

\begin{figure}[htb]
		\center \includegraphics[width=0.43\textwidth]
{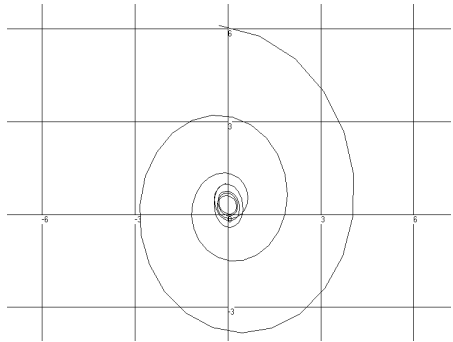} \caption[Relation between $E_x$ and $B_z$]{Relation between the fields shown in \fig{fig_wqpmf7} \ (${\hat{p}=\sqrt{\alpha},}\ \ {t\in [0,160]}$). Abscissa${: B_z(t,0,1,0)}$, \ \ Ordinate${: E_x(t,0,0,0)}$.}
	\label{fig_wqpmf8}
\end{figure}

\begin{figure}[htb]
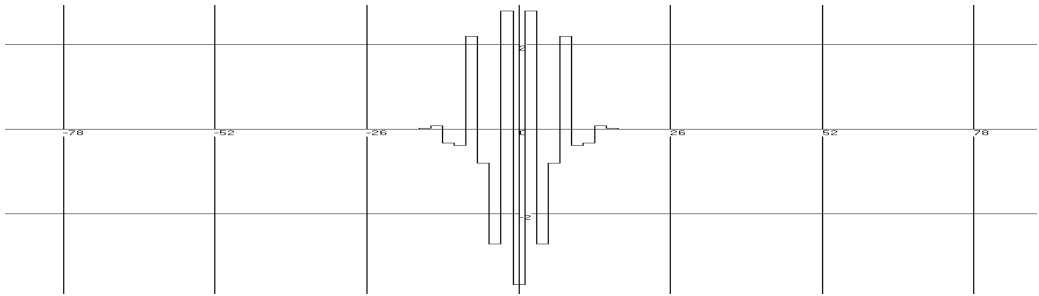
 \graic
{wqpmf9} 	\caption[Local distribution of $E_x$ after $150$ iterations]{Local distribution of $E_x$ after $150$ iterations of  (\ref{eqAlgExpplicit}) in case of ${\hat{p}=\sqrt{\alpha}}$. Abscissa${: x}$, \ \ Ordinate${: E_x(150,x,0,0)}$.}
	\label{fig_wqpmf9}
\end{figure}

We see in \fig{fig_wqpmf4}, \ref{fig_wqpmf5} and \ref{fig_wqpmf6} that  $E_x(t,0,0,0)$ begins to diverge considerably from the $6$th maximum on. \fig{fig_wqpmf7} and \ref{fig_wqpmf8} show the development of $E_x(t,0,0,0)$ together with $B_z(t,0,1,0)$ and \fig{fig_wqpmf9} the local distribution of $E_x$ after $150$ iterations of (\ref{eqAlgExpplicit}).

\begin{figure}[htb] \graic 
{wqpmf6} \caption[Long term development in case of ${\hat{p}=1/8}$]{Long term development in case of ${\hat{p}=1/8}$. Abscissa${: t}$, \ \ Ordinate${: E_x(t,0,0,0)}$.}
	\label{fig_wqpmf6}
\end{figure}

\clearpage

\section{Comment}
It was not surprising that after many iterations of (\ref{eqAlgExpplicit}) a wave packet results which is concentrated round the origin (\fig{fig_wqpmf9}). \emph{As long as} there is one and the same $\hat{p}$ for all directions, no direction is preferred. However, it was surprising for us that beginning with the $6$th maximum there must be a dominating positive feedback from the surrounding which causes divergence of $E_x(t,0,0,0)$ (and of all other field components in the origin). We noticed this also for $\hat{p}<1/16$. Further numerical tests indicated in case of $\hat{p}=\sqrt{\alpha}$ and $t\in[200,1000]$ an exponential increase of the wave maxima with an average growth factor of about $1.04$ per step $t\to t+1$. Of course it is possible to force a stable development and conservation of energy by simple normalization, e.g. by division of the right sides of (\ref{eqAlgExpplicit}) by an appropriate function $U(t)\ge 1$, but we would like to have a good justification for doing that. Remembering \ref{chapCarefulInterpretation} we guess that more advanced changes of (\ref{eqAlgExpplicit}) are necessary. Although the equations (\ref{eqAlgExpplicit}) are not exact, they can be suitable for approximative calculation of the initial discrete short term development (see \ref{chapShortTermDevelopment}) because they are directly derived from the Maxwell equations. We have done this without any usage of our knowledge of the fine structure constant, and it is noteworthy that the first maxima of $E_x(t,0,0,0)$ are nearly equal as shown in \fig{fig_wqpmf2}, if the in (\ref{eqDefp}) defined absolute coupling factor $\hat{p}$ is the root of the fine structure constant.

\section{Conclusions}
The (vacuum) Maxwell equations cannot describe physical reality exactly, like all differential equations \cite{or1}. Straightforward conversion of the vacuum Maxwell equations into finite-difference equations leads to a system of equations (\ref{eqAlgExpplicit}) which is not exact, too. But it is unambiguous, provides additional combinatorial details and can be suitable for approximative calculation of the initial discrete short term development of the electromagnetic fields. The absolute coupling factor $\hat{p}$ is a new variable which \emph{automatically} arises during formation of the finite differences. In case of $\hat{p}=\sqrt{\alpha}$ a short term wave-like development with nearly equal initial maxima results. If this is not coincidental, the definition of $\hat{p}$ (cf. (\ref{eqDefp}) together with \ref{chapmindiffasunits}) suggests a new interpretation of the fine structure constant $\alpha$ and (\ref{eqAlgExpplicit}) can serve as starting point for improvements and further combinatorial studies.

\end{document}